\documentstyle[12pt]{article}
\begin{document}
\centerline{\bf{\large{Gravitational field of a stationary circular cosmic
 string loop}}}
\vspace{5mm}
\centerline{A.A.Sen{\footnote{e-mail: anjan@juphys.ernet.in}} and N.Banerjee}
\vspace{5mm}
\begin{center}
Relativity and Cosmology Research Center\\
Department of Physics, Jadavpur University\\
Calcutta 700032, India
\end{center}
\vspace{5mm}
{\bf{Abstract}}

Gravitational field of a stationary circular cosmic string loop has been
studied in the context of full nonlinear Einstein's theory of gravity. It
has been assumed that the radial and tangential stresses of the loop are
equal to the energy density of the string loop. An exact solution for the 
system has been presented which has a singularity at a finite distance
from the axis,but is regular for any other distances from the axis of the 
loop.\\

 \vspace{5mm}
 PACS NO: 04.20Jb, 98.80.Cq \\
 \vspace{2mm}
 \newpage

Phase transition in the early universe might have produced some topological defects
\cite{R1}. One of these defects, cosmic strings, have attracted a lot of interests
amongst cosmologists for various reasons\cite{R2}. Cosmic strings are capable of
producing observational effects such as the double images of quasars and also
they are considered to be possible candidates as seeds for the galaxy formation.
Moreover, strings are believed to sustain the cosmological evolutions unlike other
defects.

In some gauge models, strings do not have any end, and thus they are either of 
infinite extent form or closed circular loops. One of the most notable features
of the gravitational field of a straight infinite cosmic string is the presence  
of an "angular deficit" in an otherwise Minkowskian spacetime having a magnitude
related to the linear energy density $\mu$ of the string by the equation
$\delta\phi=8\pi G\mu$. In fact, this angular deficit plays the key role for 
the production of the double images of a quasars\cite{R3}-\cite{R6}.

The deficit angle model is widely believed to be a good approximation for describing
a spacetime exterior to the string core. Frolov, Israel and Unruh (FIU)\cite{R7}
used this approximation for studying a closed circular cosmic string loop at a
moment of time symmetry. With the help of the initial value formulation\cite{R8},
they produced a family of momentarily stationary circular cosmic strings, which are
considered as thin loops either at the time of formation or at the turning point
of expansion and collapse. An important assumption in this work is that all points
on the circular string be conical singularities with angular deficit equal to 
that of a infinite straight string of equal linear energy density. Hughes,
McManus, and Vandyck\cite{R9} investigated further the problem of angular 
deficit of circular string. They considered a weak field stationary solutions
of Einstein's field equations for a thin circular string and also established
that radial stress should be introduced to support the string loop against
possible gravitational collapse. They determined the form of the radial stress
from the stress energy conservation relations. The main result of their study
is that under weak field assumption, circular string produces the same angular
deficit as straight infinitely long string having the same linear energy density.
In another work, McManus and Vandyck\cite{R10} considered a string loop with
a rotation of the loop which provides the necessary centrifugal reaction
partially or fully in order to avoid the possible gravitational collapse.
 
In the present work, we have assumed the form of the energy momentum tensor for
circular string as that proposed by Hughes et.al\cite{R9}. But unlike the work
of Hughes et.al, we have taken the nonzero components of $T^{\mu}_{\nu}$ to be
function of both r and z. With some simplified assumptions, we have been able to
produce a set of exact analytical solutions of the full nonlinear Einstein's
equations for a circular string. The spacetime, we have obtained, have a 
singularity at a finite distance from the axis, but is regular for any 
other distances from the axis of the loop.

A circular cosmic string loop produces an axially symmetric gravitational
field, and the most general static metric of this type may be written as
\cite{R11}
\begin{equation}
ds^2=-e^{2\nu}dt^2+e^{2(\xi-\nu)}r^2 d\phi^2+e^{2(\eta-\nu)}(dr^2+dz^2)
\end{equation}
where the three functions $\nu,\xi,\eta$ are functions of $r$ and $z$ and 
($t,\phi,r,z$) denotes the cylindrical coordinates.

We have taken the form of the energy momentum tensor for the circular string
loop as prescribed by Hughes et.al\cite{R9}:
\begin{eqnarray}
T^t_t=-\alpha(r,z)\nonumber\\
T^\phi_\phi=-\beta(r,z)\\
T^r_r=-\Delta(r,z)\nonumber
\end{eqnarray}
and all other components of $T^\mu_\nu$ are zero.
For a circular string $T^\phi_\phi$ plays the same role as that of the 
longitudinal stress $T^z_z$ for a straight string. $T^r_r$ is the external 
radial stress, required to support the loop against collapse.

In order to solve the full nonlinear Einstein's equations 
($G^\mu_\nu=8\pi T^\mu_\nu$)
(we have assumed $G=c=1$) for the system given by (1) and (2), we make an
assumption that all the nonzero components of $T^\mu_\nu$ are equal, i.e.
\begin{equation}
T^t_t=T^\phi_\phi=T^r_r= -\sigma(r,z)
\end{equation}
With this assumption and with the line element (1), the Einstein's equations
become:
$$
\ddot{\xi}-2\ddot{\nu}+\ddot{\eta}+{2\over{r}}(\dot{\xi}-\dot{\nu})+ (\dot{\xi}-
\dot{\nu})^2+\xi^{''}-2\nu^{''}+\eta^{''}+(\xi^{'}-\nu^{'})(\xi^{'}-\eta^{'})
=8\pi\sigma e^{2(\eta-\nu)}
\eqno{(4a)}$$
$$
\ddot{\eta}+\dot{\nu}^{2}+\eta^{''}+\nu^{'2}=8\pi\sigma e^{2(\eta-\nu)} 
\eqno{(4b)}
$$
$$
{\dot{\eta}\over{r}}+\dot{\eta}\dot{\xi}+\dot{\nu}^{2}+\xi^{''}+\xi^{'2}
+\nu^{'2}-\xi^{'}\eta^{'}=8\pi\sigma e^{2(\eta-\nu)} 
\eqno{(4c)}
$$
$$
\ddot{\xi}+{1\over{r}}(2\dot{\xi}-\dot{\eta})+\dot{\xi}^{2}+\dot{\nu}^{2}-
2\dot{\xi}\dot{\nu}+\xi^{'}\eta^{'}-\nu^{'2}=0 
\eqno{(4d)}
$$
$$
{1\over{r}}(\xi^{'}-\eta^{'})+\dot{\xi}(\eta^{'}-\xi^{'})+\xi^{'}\dot{\eta}-
2\dot{\nu}\nu^{'}=0
\eqno{(4e)}
$$

In the above equations an overhead dot represents differentiation w.r.t $r$ and
prime represents differentiation w.r.t $z$.

The conservation equations ($T^\mu_{\nu;\mu}=0$) give the relations
$$
\dot{\sigma}+\sigma(\dot{\eta}-\dot{\nu})=0
\eqno{(5)}
$$
and
$$
\sigma(\xi^{'}+\eta^{'}-\nu^{'})=0
\eqno{(6)}
$$

We see immediately from (6) that, in order to have the energy density $\sigma$
to be nonzero, we should have,
$$
\xi^{'}=\nu^{'}-\eta^{'}.
\eqno{(7)}
$$
Equation (5) can be integrated to give
$$
\sigma=C(z)e^{(\nu-\eta)}
\eqno{(8)}
$$
where $C(z)$ is an arbitary function of $z$.

Now we make the simplifying assumption that the metric coefficients are 
separable functions of their arguments i.e. $\xi=\xi_{1}(r)+\xi_{2}(z)$,
$\nu=\nu_{1}(r)+\nu_{2}(z)$ and $\eta=\eta_{1}(r)+\eta_{2}(z)$. Equations (8) 
then shows that $\sigma$ will be also separable function of its arguments.

In what follows we will find the possible solutions of the full nonlinear
Einstein's equations given by (4a)-(4e) with these assumptions.\\
Equations (4a)-(4e) can now be written as,
$$
\ddot{\xi_{1}}-2\ddot{\nu_{1}}+\ddot{\eta_{1}}+{2\over{r}}(\dot{\xi_{1}}-
\dot{\nu_{1}})+ (\dot{\xi_{1}}-\dot{\nu_{1}})^2
+\xi_{2}^{''}-2\nu_{2}^{''}+\eta_{2}^{''}+(\xi_{2}^{'}-\nu_{2}^{'})(\xi_{2}^{'}
-\eta_{2}^{'})
=8\pi\sigma e^{2(\eta-\nu)}
\eqno{(9a)}$$
$$
\ddot{\eta_{1}}+\dot{\nu_{1}}^{2}+\eta_{2}^{''}+\nu_{2}^{'2}
=8\pi\sigma e^{2(\eta-\nu)} 
\eqno{(9b)}
$$
$$
{\dot{\eta_{1}}\over{r}}+\dot{\eta_{1}}\dot{\xi_{1}}+\dot{\nu_{1}}^{2}+
\xi_{2}^{''}+\xi_{2}^{'2}
+\nu_{2}^{'2}-\xi_{2}^{'}\eta_{2}^{'}=8\pi\sigma e^{2(\eta-\nu)} 
\eqno{(9c)}
$$
$$
\ddot{\xi_{1}}+{1\over{r}}(2\dot{\xi_{1}}-\dot{\eta_{1}})+\dot{\xi_{1}}^{2}+
\dot{\nu_{1}}^{2}-
\dot{\xi_{1}}\dot{\eta_{1}}+\xi_{2}^{'}\eta_{2}^{'}-\nu_{2}^{'2}
=0 
\eqno{(9d)}
$$
$$
{1\over{r}}(\xi_{2}^{'}-\eta_{2}^{'})+\dot{\xi_{1}}(\eta_{2}^{'}-\xi_{2}^{'})
+\xi_{2}^{'}\dot{\eta_{1}}-
2\dot{\nu_{1}}\nu_{2}^{'}=0
\eqno{(9e)}
$$

As the right hand side of both equations (9a) and (9b) are equal, one can equate 
the $r$ part and $z$ part of these equations.

By equating $r$ part of (9a) and (9b), one can get after some simple calculations
$$
\dot{\xi_{1}}+{1\over{r}}=2\dot{\nu_{1}}
\eqno{(10)}
$$
With the assumption of separability of the metric coefficients, 
the z part of equation (7) reads like
$$
\xi_{2}^{'}=\nu_{2}^{'}-\eta_{2}^{'}
\eqno{(11)}
$$
With (11) and (10), one can have from (9e), the relation,
$$
\xi_{2}^{'}(4\dot{\nu_{1}}-\dot{\eta_{1}})=0
\eqno{(12)}
$$
From eqn(12), it is easy to calculate that either $\xi_{2}^{'}=0$ or
$4\dot{\nu_{1}}=\dot{\eta_{1}}$.

In this work, we present the solutions  where $\xi_{2}^{'}=0$ and
$4\dot{\nu_{1}}\neq\dot{\eta_{1}}$.\\
Therefore,
$$
e^{2\xi_{2}}=C_{0},
\eqno{(13)}
$$
$C_{0}$ being a constant.\\
From eqn(11), with the help of (13) one gets,
$$
\nu_{2}^{'}=\eta_{2}^{'}
\eqno{(14)}
$$
With eqn(13), eqn(9d) becomes,
$$
\ddot{\xi_{1}}+{1\over{r}}(2\dot{\xi_{1}}-\dot{\eta_{1}})+\dot{\xi_{1}}^{2}+
\dot{\nu_{1}}^{2}-
\dot{\xi_{1}}\dot{\eta_{1}}=\nu_{2}^{'2}=K
\eqno{(15)}
$$
where $K$ is the constant of separation.\\
With (10),(13),(14) and (15), (9a) and  (9b) reduce to a single equation,
$$
\ddot{\eta_{1}}+\dot{\nu_{1}}^{2}+K
=8\pi\sigma e^{2(\eta_{1}-\nu_{1})} 
\eqno{(16)}
$$
and eqn(9c) becomes
$$
{\dot{\eta_{1}}\over{r}}+\dot{\eta_{1}}\dot{\xi_{1}}+\dot{\nu_{1}}^{2}+K
=8\pi\sigma e^{2(\eta_{1}-\nu_{1})} .
\eqno{(17)}
$$
From the $z$ part of eqn(15) and using (14) we have
$$
e^{\eta_{2}}=e^{\nu_{2}}=A_{0}e^{\sqrt{K}z}
\eqno{(18)}
$$
where $A_{0}$ is an arbitrary constant of integration.\\
Combining eqn(16) and (17), we get
$$
\ddot{\eta_{1}}-{\dot{\eta_{1}\over{r}}}-\dot{\eta_{1}}\dot{\xi_{1}}=0
$$
which, with the use of eqn(10) becomes
$$
\ddot{\eta_{1}}-2\dot{\nu_{1}}\dot{\eta_{1}}=0,
\eqno{(19)}
$$
for which $\dot{\eta_{1}}=0$ is a solution which is used in our subsequent 
calculations. The other case is ${\ddot{\eta_{1}}\over{\dot{\eta_{1}}}}=
2\dot{\nu_{1}}$, and this is not considered in the present work.\\
Hence for our case
$$
e^{2\eta_{1}}=constant=m.
\eqno{(20)}
$$
Using (20) and (10) in the $r$ part of eqn(15), we get on integration
$$
e^{\nu_{1}}=e^{a(r-r_{0})}[1-e^{-5a(r-r_{0})}]^{2/5}
\eqno{(21)}
$$
where $a^{2}=K/5$ and $r_{0}$ is an arbitary integration constant.\\
Eqn(10) and (21) together yield the solution of $\xi_{1}$ as
$$
e^{\xi_{1}}={1\over{r}}e^{2a(r-r_{0})}[1-e^{-5a(r-r_{0})}]^{4/5}.
\eqno{(22)}
$$
where a constant of integration has been ignored without any loss of 
generality.\\

Hence the complete line element for a circular cosmic string loop with radial 
and tangential stress equal to the energy density, is given by
$$
ds^2=e^{2a(r-r_{0})}[1-e^{-5a(r-r_{0})}]^{4/5}(-A_{0}^{2}e^{2nz}dt^2
+C_{0}A_{0}^{-2}e^{-2nz}d\phi^{2})$$
$$
\hspace{6mm}+me^{-2a(r-r_{0})}[1-e^{-5a(r-r_{0})}]^{-4/5}(dr^2+dz^2) 
\eqno{(23)}
$$
where $n=\sqrt{K}$ and $a^{2}=K/5$.\\
One can calculate the energy density $\sigma$ from either (16) or (17) which 
becomes,
$$
\sigma={K\over{8\pi m}}e^{2a(r-r_{0})}[1-e^{-5a(r-r_{0})}]^{4/5}
[{{(e^{-5a(r-r_{0})}+1)^{2}}\over{5(e^{-5a(r-r_{0})}-1)^{2}}}+1]
\eqno{(24)}
$$
One can see from (24) that $\sigma$ is regular everywhere except at 
$r=r_{0}$ and hence $r=r_{0}$ is identified to correspond to the radius
of the loop.

We have a circular ring at $r=r_{0}$, $z=0$ momentarily at rest at time $t=0$.
At this point it is convenient to pass from the coordinates $r,z$ to toroidal
coordinates $\alpha, \psi$ defined by

$$
r= r_{0} N^{-2} sinh\alpha, z=r_{0} N^{-2} sin\psi
\eqno{(25a)}
$$
with 
$$
N(\alpha,\psi) = (cosh\alpha - cos\psi)^{1/2} \:(0\leq\alpha\leq\infty,
-\pi\leq\psi\leq\pi)
\eqno{(25b)}
$$          

We note that the circular loop $r=r_{0},z=0$ corrsponds to the limiting torus
$\alpha\rightarrow\infty$.

The metric (1) is now transformed to toroidal coordinates as
$$
ds^2=-e^{2\nu}dt^2+e^{2(\xi-\nu)}r^2 d\phi^2+{r_{0}}^{2}N^{-4}e^{2(\eta-\nu)}
(d\alpha^2+d\psi^2) 
\eqno{(26)}
$$
where $\nu,\xi,\eta,r$ are functions of $\alpha$ and $\psi$.

To check the metric (26) for elementary flatness on the loop i.e. 
for $\alpha\rightarrow\infty$ \cite{R7,R9} we must examine circle 
$\alpha=\alpha_{0}$ at constant $t,\phi$ around the ring and calculate the
ratio of the proper perimeter to proper radius in the proper radius tends to
zero.

For $t,\phi$ constant, the metric (26) using (18) and (20) becomes
$$
ds^2 = {r_{0}}^{2} m N^{-4} e^{-2\nu_{1}}(d\alpha^{2} + d\psi^{2})
\eqno{(27)}
$$
To be elementary flat on the loop like Hughes et.al \cite{R9} 
$
F = \left | \lim_{\alpha_{0}\rightarrow\infty}{\int_{-\pi}^{\pi}d\psi N^{-2}
e^{-\nu_{1}}|_{\alpha=\alpha_{0}}\over{\int_{\alpha_{0}}^{\infty}N^{-2}
e^{-\nu_{1}}d\alpha}} \right | 
$ should be a constant.

Using (21) and (25) one can calculate that 
$$
e^{-\nu_{1}}\rightarrow b_{0}e^{4\alpha/5}
\eqno{(28)}
$$
for $\alpha\rightarrow \infty$ where $b_{0} = (-10ar_{0})^{-2/5}$.\\
From (25b)
$$
N^{-2} \rightarrow 2e^{-\alpha}
\eqno{(29)}
$$
for $\alpha\rightarrow\infty$.

Using (28) and (29) one can see that 
$$
F\rightarrow 2\pi/5
\eqno{(30)}
$$
for $\alpha \rightarrow\infty $. And hence one can calculate the angular 
deficit near the loop to be $d\psi = 8\pi/5$.

In Conclusion we have presented the exact line element for a circular loop
in the context of a full set of nonlinear Einstein's theory of gravity. The
solutions presented here are, however, not the most general ones, as we have
assumed that both the radial and the tangential stresses are equal to the 
energy density of the loop and also assumed that the metric components are 
separable as products of functions of $r$ and $z$, but this is perhaps the 
first attempts for obtaining the solutions of the full nonlinear Einstein's 
equations  for a circular cosmic string loop. The earlier solution for the 
spacetime for a cicular string loop , given by Hughes et.al \cite{R9} are 
based on a linearized approximation of the field equations. In the present 
solution the energy density for the loop blows at a certain $r=r_{0}$ which
evidently corressponds to the site of the string loop.

\vspace{5mm}
The authors are grateful to the referee for his valuable comments and 
suggestions and also for pointing out some serious typographical errors.
This work is partially supported by the D.S.T., Govt.of India. One of the 
author (A.A.S) is thankful to the University Grants Commission, India 
for financial support.

\end{document}